\documentclass[12pt,preprint]{aastex}

\newcommand\chbobs{$C({\rm H}\beta)^{obs}$}
\newcommand\chbcol{$C({\rm H}\beta)^{col}$}
\newcommand\chbdust{$C({\rm H}\beta)^{dust}$}
\newcommand{\grsim}{\mathrel{\hbox{\rlap{\hbox{\lower4pt\hbox{$\sim$}}}\hbox{$>$}}}}

\shorttitle{Primordial Helium}
\shortauthors{Peimbert et al.}

\begin{document}

\title{Revised Primordial Helium Abundance Based on New Atomic Data}

\author{Manuel Peimbert}
\affil{Instituto de Astronom\'\i a, Universidad Nacional Aut\'onoma de M\'exico, 
Apdo. Postal 70-264, M\'exico 04510 D.F., Mexico}
\email{peimbert@astroscu.unam.mx}

\author{Valentina Luridiana} 
\affil{Instituto de
Astrof\'{\i}sica de Andaluc\'{\i}a (CSIC), Camino Bajo de Hu\'etor 50,
18008 Granada, Spain}
\email{vale@iaa.es}

\and
\author {Antonio Peimbert}
\affil {Instituto de Astronom\'\i a, Universidad Nacional Aut\'onoma de M\'exico,
Apdo. Postal 70-264, M\'exico 04510 D.F., Mexico} 
\email{antonio@astroscu.unam.mx}

\begin{abstract}
We have derived a primordial helium abundance of $Y_p = 0.2477 \pm 0.0029$,
based on new atomic physics computations of the
recombination coefficients of He{\sc~i} and of the collisional excitation of the H{\sc~i} Balmer
lines together with
observations and photoionization models of metal-poor extragalactic 
H{\sc~ii} regions. The new atomic data increase our previous 
determination of $Y_p$ by 0.0086, a very significant amount. 
By combining our $Y_p$ result with the predictions made by the standard Big Bang nucleosynthesis
model, we find a baryon-to-photon ratio, $\eta$, in excellent agreement both with the $\eta$ value 
derived by the primordial deuterium abundance value observed in damped Lyman-$\alpha$ systems
and with the one obtained from the WMAP observations.
\end {abstract}


\section{Introduction}\label{sec:intro}

This is the third paper of a series on the determination of the primordial helium abundance 
by unit mass, $Y_p$. In Paper I \citep{pea02} we studied the effect of temperature
variations on the determination of $Y_p$, and in Paper II \citep{lal03} we studied 
the effect of collisional excitation of the Balmer lines in the determination of $Y_p$.

The determination of $Y_p$ is important for the study of cosmology and the 
chemical evolution of galaxies. In particular, $Y_p$ can be used to test the
standard Big Bang nucleosynthesis (SBBN) scenario by setting strong constraints on: 
a) the number of neutrino families, $N_\nu$; b)
the variation of the neutron lifetime and of the neutron-proton mass difference,
these constraints can be translated into constraints on the time variation of the 
Yukawa couplings and the fine structure constant;
c) the variation of the constant of gravity, $G$; d) the presence of vacuum energy during
BBN; and e) the presence of decaying particles during BBN, which could have affected the production of the
light elements \citep[e.g.][and references therein]{cyb05,coc06}.
The accuracy in $Y_p$ needed to reach these goals extends to the third decimal 
place \citep[e.g.][]{pei03,lur03,ste06a}. $Y_p$ determinations are affected 
by at least thirteen sources of error \citep[see Section~\ref{sec:abund} and the 
review by][]{pei03}. In Paper II we obtained a determination of $Y_p$ in which most 
of these sources of error were taken into account.
In this paper we improve our previous $Y_p$ determination, considering new atomic data
and improved stellar population synthesis models, which are likely to further reduce
the error affecting our computation. Specifically, 
four of the main sources of error are reduced by means of the
use of: a) the recent He{\sc~i} recombination
coefficients by \citet{por05,por07}, b) the H{\sc~i} collisional excitation
coefficients by \citet{and00,and02}, and c) the correction for
underlying absorption for lines redward of 5000 \AA, based on the population 
synthesis models by \citet{gon05} and the observations by \citet{leo98}.

$Y_p$ is determined by means of an extrapolation to $Z=0$ of the $Y$ values
of a sample of objects. Here, the usual
normalization $X + Y + Z = 1$ is used, where $X$, $Y$, and $Z$ are the abundances
per unit mass of hydrogen, helium, and the rest of the elements respectively.
The extrapolation is traditionally done in the $Y,Z$ space by assuming a $\Delta Y/\Delta Z$ slope \citep{pei74}. More recently it has become common
to use $\Delta Y/\Delta O$, where $O$ is the oxygen
abundance per unit mass, since the $O$ value is easier to determine.

In Section~\ref{sec:collisions} we discuss the collisional excitation of the 
Balmer lines and apply the corresponding correction to our H{\sc~ii} regions 
sample. In Section~\ref{sec:abund} we correct the observed line intensities 
by extinction and
underlying absorption, also in this section we determine the $Y$ values
for each of the objects of the sample, and combining them with a 
$\Delta Y/\Delta O$ relationship we derive the new $Y_p$ value. In
this section we present the error budget of our determination and
a discussion is made where we distinguish between systematic and
statistical effects.
In Section~\ref{sec:comparison} we compare our determination with those 
by other authors. In Section~\ref{sec:cosmology} we discuss the cosmological 
implications of our new $Y_p$ value, and in Sections~\ref{sec:discussion} 
and \ref{sec:conclusions} we present the discussion and our conclusions. A 
preliminary account of some of the results
included in this paper was presented elsewhere \citep{pei07}.

\section{Collisional enhancement of hydrogen Balmer lines}\label{sec:collisions}

In high-temperature H{\sc~ii} regions, the hydrogen Balmer lines can be excited by collisions of neutral atoms with free electrons. This effect is generally small, usually contributing less than a few percent of the intensity 
of H$\alpha$ and H$\beta$ or less, which for most applications can be neglected.
However, collisional enhancement must be taken into account in the determination of $Y_p$ for at least two reasons. First, for this task the helium abundance by unit mass, $Y$, in individual H{\sc~ii} regions must be known with the highest attainable precision. Second, the objects where collisions are more important are those with the highest $T_e$ and, hence, the lowest metallicities; consequently, they bear a major weight in the extrapolation to $Z=0$ of the relation between $Y$ and $Z$.

In Paper II we used photoionization models calculated with the photoionization code Cloudy to estimate the collisional contribution to the Balmer lines in five low-metallicity objects. For three objects (Haro~29, I~Zw~18, and SBS~0335-052) we computed tailored models using version 94.00 of Cloudy, last described by \citet{fer98}; for the remaining two objects(NGC~2363 and NGC~346) the models were retrieved from previously published works 
\citep[][respectively]{lur99,rel02}, and the collisional rate in them was estimated from the
difference between the total and the case B H$\beta$ emission predicted
by the models. In the present work, we present improved estimates, which differ from the previous ones in several aspects: a) The models for NGC~2363 and NGC~346 have been recalculated based on the input parameters described in the original papers, so that the collisional contribution could be explicitly computed rather than simply estimated; b) the collisional rates have been updated; c) the radiative cascade following collisional excitation is now computed self-consistently, and d) the model fitting follows a different philosophy with respect to Paper II. Points b), c) and d) will be explained in detail in the following sections. A further difference with respect to Paper II is that, while the collisional luminosities of Paper II models had been computed simultaneously with the models themselves by modifying a Cloudy's routine, those of the present paper have been computed by an external program, to which Cloudy's output (i.e., the ionization and temperature structures) is fed. The difference between the two procedures is, of course, irrelevant from the point of view of results, except for the change in the collisional atomic data used in either case, which will be discussed in Section~\ref{sec:updated_omegas}.

\subsection{Updated collisional rates}\label{sec:updated_omegas}

In previous works, it has been estimated that $I({\rm H}\beta)_{col}$/$I({\rm H}\alpha)_{col}$ is about 1/10 \citep{dav85,sta01,lal03}, and it has been predicted that this differential enhancement would increase the observed Balmer decrement, mimicking a larger interstellar reddening. The amount of extra reddening due to this mechanism is expressed by the collisional reddening coefficient \chbcol:
\begin{eqnarray}
C({\rm H}\beta)^{\rm col} & = &\frac{{\rm Log}(I({\rm H}\alpha)^{\rm tot}/I({\rm H}\beta)^{\rm tot})
- {\rm Log}(I({\rm H}\alpha)^{\rm rec}/I({\rm H}\beta)^{\rm rec})}{-f({\rm H}\alpha)} = \nonumber \\
&=&\frac{{\rm Log}\bigl((1-I({\rm H}\beta)^{\rm col}/I({\rm H}\beta)^{\rm tot})/(1-I({\rm H}\alpha)^{\rm col}/I({\rm H}\alpha)^{\rm tot})\bigr)}{-f({\rm H}\alpha)}= \nonumber \\
&\equiv&\frac{{\rm Log}\bigl((1-x_\beta)/(1-x_\alpha)\bigr)}{-f({\rm H}\alpha)},
\end{eqnarray}
where $I(\lambda)_{rec}/I(\lambda)_{tot}$ and $x_\lambda \equiv I(\lambda)_{col}/I(\lambda)_{tot}$  
are the relative contributions of recombinations and collisions, respectively, 
to the total emitted intensity in $\lambda$, and $f({\rm H}\alpha)$ is the reddening correction at $H\alpha$.
Collisional and interstellar reddening add together to yield the observed reddening:
\begin{equation}
C({\rm H}\beta)^{\rm obs} = C({\rm H}\beta)^{\rm dust} + C({\rm H}\beta)^{\rm col}, 
\end{equation}
(see Paper II), so that a good estimate of \chbcol\ is needed to properly derive
\chbdust\ from the observed reddening.

In the present work we have used updated collisional coefficients \citep{and00,and02} to revise our previous estimates of $x_\alpha$ and $x_\beta$ and the derived value of $C({\rm H}\beta)^{\rm col}$. Two major features of the new data deserve to be noted here in comparison to those used in Paper II \citep{cal94,vri80}\footnotetext{The reference to \citet{and00} and \citet{and02} contained in Paper II was erroneous, as the models had been computed with an earlier version of Cloudy.}:

\begin{enumerate}

\item All the collisional coefficients $\Omega(1, n)$, to which collisional rates are directly proportional, are larger than the corresponding values by \cite{cal94} and \cite{vri80} in the temperature range of interest.

\item The increase of $\Omega(1,4)$ is larger than that of $\Omega(1,3)$ by a factor of 2.5 approximately. 

\end{enumerate}

Point 1 above implies larger collisional contributions to all Balmer lines than previously estimated. 
Point 2 implies that $I({\rm H}\beta)_{col}$/$I({\rm H}\alpha)_{col}$ also increases by (roughly) the same factor of 2.5 as 
$\Omega(1,4)$/$\Omega(1,3)$. This is because, although collisional excitations to any level with $n\ge3$ might eventually produce an H$\alpha$ photon, excitations to $n=3$ largely dominate and are responsible for more than 80\% of the collisional H$\alpha$ emission in our objects; analogously, excitations to $n=4$ dominate the collisional H$\beta$ emission, yielding more than 90\% of it. Therefore, the predicted $I({\rm H}\beta)_{col}$/$I({\rm H}\alpha)_{col}$ closely follows the $\Omega(1,4)$/$\Omega(1,3)$ ratio: with the data by \citet{and00, and02} it is now predicted to be larger than 1/4 at $T_e = 20000$ K and approaching 1/3 at larger temperatures, making it almost indistinguishable from the normal Balmer decrement produced by recombinations. As a result, with the updated collisional coefficients \chbcol\ 
is predicted to be much smaller than previously estimated, and a larger fraction of the observed reddening can be attributed to dust interstellar reddening.

\subsection{Improved radiative cascade}\label{sec:radiative_cascade}

In addition to using the new collisional coefficients, we have now improved our scheme for the calculation of the radiative cascade following a collisional excitation. The collisional data used in Paper II were available only as
level-averaged coefficients $\Omega(n)$, whereas the new coefficients are available as $\Omega(n,l)$.
Therefore, an ad-hoc assumption had to be made in Paper II regarding how electrons excited by collisions distribute themselves among the sublevels $(n,l)$ of a given level $n$; this assumption, in turn, determined the radiative spectrum following an upward collision.
In the present work, in contrast, knowledge of the $\Omega(n,l)$ makes it straightforward to calculate the radiative cascade following upward collisions, and no assumptions are required (for low values of $n$, $l$-mixing is completely negligible at the densities of our objects, $N_e\lesssim 1000$ cm$^{-3}$). 

\subsection{Model fitting}\label{sec:model_fitting}

Because the collisional enhancement of Balmer lines depends on the interplay between the abundance of H$^0$ and the temperature structure of the region, good estimates can only be given with tailored photoionization models that simultaneously constrain the temperature and the ionization structure. In Paper II, we computed several tailored photoionization models for each of the three high-$T_e$ H{\sc~ii} regions SBS 0335-052, I~Zw~18 and Haro~29, and estimated the collisional enhancement of H$\alpha$ and H$\beta$ by selecting those models
that properly fit the observed [O {\sc III}] temperature, without bothering about the fitting of $T$~[O {\sc II}]. The rationale of that choice was that, because of the Boltzmann factor in the expression for the collisional rate, the temperature dependence of collisional rates is very strong: as a consequence, we thought that an accurate estimate could only be granted by a good fit of the hottest region of the nebula, which in this range of metallicities is the inner one and is characterized by the [O {\sc III}] temperature.

\begin{figure}
\plotone{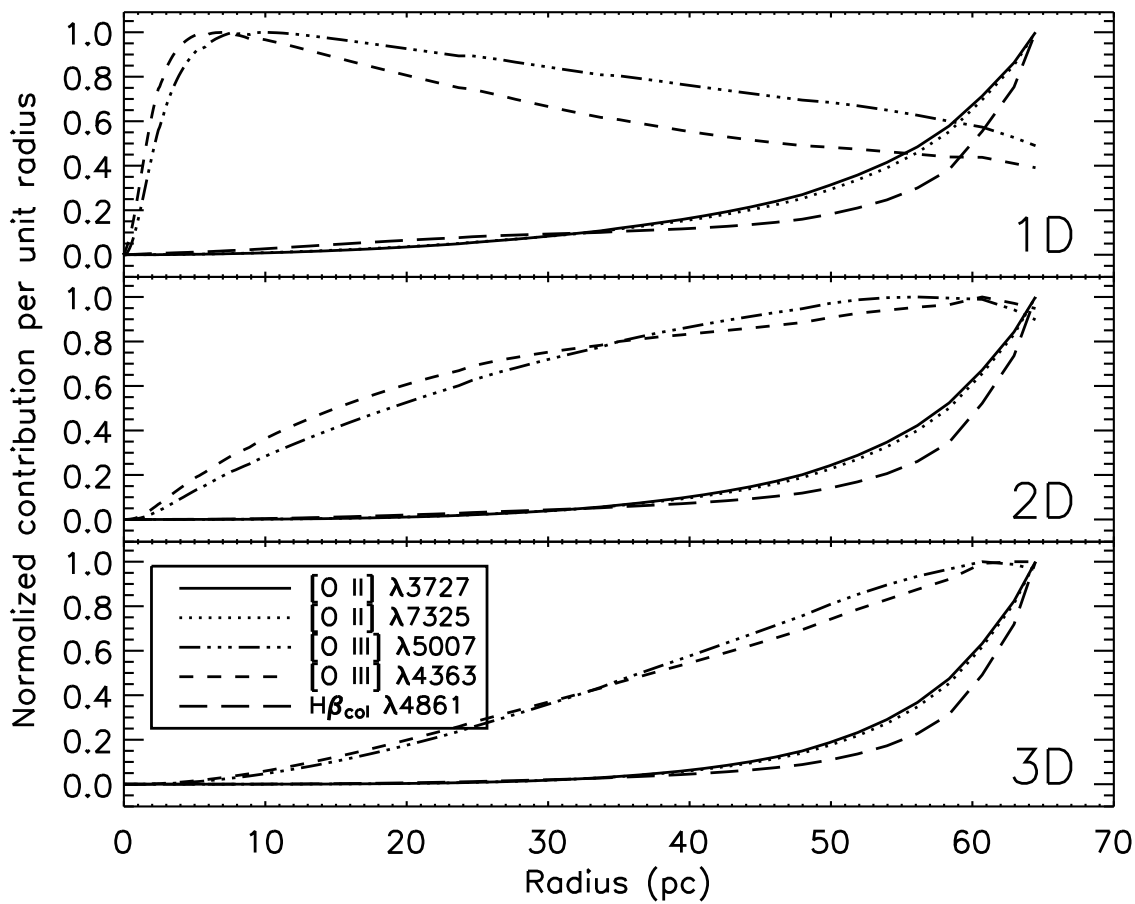}
\caption{Normalized contribution per unit radius to the emission in the main [O~{\sc II}] and [O~{\sc III}] lines 
and in $\rm{H}\beta_{col}$ in a model for NGC~346, showing that $\rm{H}\beta_{col}$ preferentially forms in the [O~{\sc II}] zone. The values are normalized to the maximum value in each line. The three panels correspond to different integrations, simulating a beam, a narrow slit, and a slit covering the whole object respectively.
\label{fig:collisions}}
\end{figure}

However, further calculations have shown that, in most cases, the dominant contribution to the collisional excitation of hydrogen comes from the outer zones. This happens because the outward increase in the fraction of neutral hydrogen outweighs the decrease in temperature, exceptions to this behavior will be mentioned in Section~\ref{sec:discussion} and extensively discussed in a forthcoming paper (Luridiana, in preparation). Since the outer zones are best characterized by the [O{\sc~II}] temperature, the collisional contribution to Balmer lines is now computed based on models that correctly fit the [O {\sc II}] rather than the [O {\sc III}] temperature.
An example of this behavior is given in Fig.~\ref{fig:collisions}, which shows, for one of our model nebulae, the relative contribution of each layer to the emission in the [O{\sc~II}] and [O {\sc~III}] lines and in $\rm{H}\beta_{col}$. The three panels correspond to different integrations, simulating a beam (i.e., a point-like slit), a narrow slit, and a slit covering the whole object respectively; depending on how the nebula is sampled by the aperture, different layers contribute differently to the overall emission. In all cases, we can see that the collisional ${\rm H}\beta$ emission closely follows the emission of the [O{\sc~ii}] lines: in other words, $\rm {H}\beta_{col}$ preferentially forms in the [O{\sc~II}] rather than in the [O {\sc~III}] zone. The effect is more pronounced in the case of a beam, which is a good approximation for the data of four of our five objects, but is still seen even in the case in which the whole nebula is sampled.

\clearpage

According to this criterion, the fitting philosophy was changed to grant agreement with the [O{\sc~ii}] rather than the  [O{\sc~iii}] temperature. The models of Haro~29 and I~Zw~18 used to this aim are the corresponding best-fit models of Paper~II, which already fitted the observed $T$[O{\sc~ii}] temperatures. For SBS~0335 we used model A of \citet{lal03}, which correctly fits the [O{\sc~ii}] temperature, instead of the hotter model B, which fits the [O{\sc~iii}] temperature and was used in Paper~II. For NGC~346 and NGC~2363 (for which in Paper II we only gave estimates based on models compiled from the literature) we computed two new models with Cloudy (version 06.02): the one for NGC~346 is based on model L4 by \citet{rel02}, and the one for NGC~2363 is based on the best-fit model for this region by \citet{lur99}. In all the models, the updated collisional rates by \citet{and00,and02} were used, and the appropriate corrections for the slit aperture used in the observations were made.

\subsection{Resulting estimates of collisional contribution to the Balmer lines}\label{sec:estimate_collisions}

The models described in Sections~\ref{sec:model_fitting}, which incorporate the changes described in sections~\ref{sec:updated_omegas} and \ref{sec:radiative_cascade}, were used to derive estimates for the collisional contributions. The results are listed in Table~\ref{tab:all_collisions}.
Based on these values, we recalculated the breakdown of \chbobs\ in terms of collisional and dust reddening for the objects of the sample. The results are listed in Table~\ref{tab:all_cHbeta}. The $x_\alpha$ and $x_\beta$ values
predicted by our model A for SBS~0335 were not published in Paper II.

The effect of using the new atomic data is an increase in the estimated collisional contribution to the
Balmer lines. This increase is partially offset by the lower temperature of the photoionization models used in this work: some of the models used in Paper II, computed to fit the 
observed $T$~[O{\sc~iii}] values, produced $T$~[O{\sc~ii}] values larger by up to 700 K 
than the observed values, while the new models correctly fit the $T$[O{\sc~ii}] 
temperature, and have correspondingly lower collisional rates.
In Table~\ref{tab:all_collisions} we list the new estimates of $x_\alpha$ and $x_\beta$ and the values of 
$T$~[O{\sc~ii}].

\clearpage
\begin{deluxetable}{lcccccc}
\tablecaption{Collisional contribution to the total Balmer intensities\tablenotemark{a}.
\label{tab:all_collisions}}
\tablewidth{0pt}
\tablehead{
\colhead{}       &
\colhead{}         &
\multicolumn{2}{c}{Paper II} &&
\multicolumn{2}{c}{This work} \\
\cline{3-4} \cline{6-7} 
\colhead{Object} &
\colhead{$T_e$ (O{\sc~II})}&
\colhead{$x_\alpha$}      &
\colhead{$x_\beta$}      &&
\colhead{$x_\alpha$}      &
\colhead{$x_\beta$}      
}
\startdata

\object{NGC 346}                                      & 12600 & 0.015 & 0.004            && 0.011 & 0.007 \\
\object{NGC 2363}                                     & 13800 & 0.030 & 0.008            && 0.037 & 0.027 \\
\object{Haro~29}                                          & 14000 & 0.028 & 0.007            && 0.033 & 0.021 \\
\object{SBS 0335-052\tablenotemark{b}}                & 15600 & 0.074 & 0.021            && 0.086 & 0.066 \\
\object{I~Zw~18}                                      & 15400 & 0.060 & 0.017            && 0.070 & 0.053 \\

\enddata
\tablenotetext{a}{$x_\lambda=I(\lambda)_{col}/I(\lambda)_{tot}$
is the ratio between the collisional and the total intensity, calculated
with the updated collisional coefficients by \cite{and00,and02} for the 
photoionization models described in \citet*{lur99}, \citet*{rel02}, and Paper II.}
\tablenotetext{b}{Predictions of model A for the centermost 1.8'' x 1'' region, corresponding to the sum of the three brightest
positions observed by Izotov et al. (1999).}
\end{deluxetable}

\begin{deluxetable}{lccccc}
\tablecaption{Observed, collisional, and collision-corrected reddening coefficients\tablenotemark{a}.
\label{tab:all_cHbeta}}
\tablewidth{0pt}
\tablehead{
\colhead{} &
\colhead{} &
\multicolumn{2}{c}{Paper II} &
\multicolumn{2}{c}{This work} \\
\cline{3-4} \cline{5-6} 
\colhead{Object}       &
\colhead{\chbobs}       &
\colhead{\chbcol}       &
\colhead{\chbdust}       &
\colhead{\chbcol}       &
\colhead{\chbdust}       
}
\startdata

\object{NGC 346}                 & $0.15\pm0.01$ & $0.02\pm0.02$ & $0.13\pm0.02$ & $0.00\pm0.00$ & $0.15\pm0.01$ \\
\object{NGC 2363}                & $0.11\pm0.02$ & $0.03\pm0.02$ & $0.08\pm0.02$ & $0.01\pm0.01$ & $0.10\pm0.02$ \\
\object{Haro~29}                     & $0.00\pm0.08$ & $0.03\pm0.02$ & $-0.03\pm0.08$ & $0.02\pm0.01$ & $-0.01\pm0.08$ \\
\object{SBS 0335-052}            & $0.24\pm0.02$ & $0.07\pm0.05$ & $0.17\pm0.06$ & $0.03\pm0.01$ & $0.21\pm0.02$ \\
\object{I~Zw~18}                 & $0.02\pm0.02$ & $0.06\pm0.02$ & $0.04\pm0.02$ & $0.02\pm0.01$ & $0.00\pm0.02$ \\

\enddata
\tablenotetext{a}{The uncertainty in \chbcol\ was calculated as in Paper II.}
\end{deluxetable}
\clearpage

\section{Abundance determinations}\label{sec:abund}

\subsection{Adopted line intensities}

In Tables~\ref{tab:lineA} and \ref{tab:lineB} we present the observed line intensity 
ratios of the He{\sc~i} lines
relative to H($\beta$),
$F(\lambda)/F({\rm H}\beta)$, for the 5 objects without
correction for underlying absorption.

\begin{deluxetable}{lr@{$\pm$}lr@{$\pm$}lcr@{$\pm$}lr@{$\pm$}lcr@{$\pm$}lr@{$\pm$}l}
\rotate
\tablecaption{Adopted He~{\sc{i}} line intensities relative to H$\beta$\tablenotemark{a}
\label{tab:lineA}}
\tablewidth{0pt}
\tablehead{
&
\multicolumn{4}{c}{NGC~346\tablenotemark{b}}&&
\multicolumn{4}{c}{NGC~2363\tablenotemark{c}}&&
\multicolumn{4}{c}{Haro~29\tablenotemark{c}}\\
\cline{2-5} \cline{7-10} 
\cline{12-15} 
\colhead{He~{\sc{i}} line}&
\multicolumn{2}{c}{$F$}&\multicolumn{2}{c}{$I$}&&
\multicolumn{2}{c}{$F$}&\multicolumn{2}{c}{$I$}&&
\multicolumn{2}{c}{$F$}&\multicolumn{2}{c}{$I$}}
\startdata
3820& 
             \multicolumn{2}{c}{...}&
		\multicolumn{2}{c}{...}&&
                        0.011&0.001&
			        0.014&0.001&&
                                       0.007&0.001&
				          0.010&0.001\\
				             
3889&
	0.1820&0.0017\tablenotemark{d}&
		0.0988&0.0019&&
			0.174&0.001\tablenotemark{d}&
				0.095&0.003&&
					0.186&0.001\tablenotemark{d}&
                                               0.097&0.001\\
4026&
	0.0171&0.0006&
		0.0203&0.0007&&
			0.015&0.001&
				0.020&0.001&&
					0.016&0.001&
                                             0.019&0.001\\
4387&
        0.0045&0.0002&
		0.0054&0.0002&&
			0.004&0.001&
				0.005&0.001&&
					0.004&0.001&
                                             0.005&0.001\\
4471&
        0.0370&0.0005&
		0.0396&0.0005&&
			0.038&0.001&
				0.042&0.001&&
					0.037&0.001&
                                             0.040&0.001\\
4922&
        0.0100&0.0002&
		0.0109&0.0002&&
			0.012&0.001&
			        0.013&0.001&&
					0.009&0.001&
                                             0.010&0.001\\
5876&
        0.1143&0.0013&
		0.1075&0.0012&&
			0.112&0.001&
				0.110&0.002&&
					0.103&0.001&
                                             0.105&0.001\\
6678&
        0.0336&0.0002&
		0.0305&0.0002&&
			0.031&0.001&
				0.030&0.001&&
					0.029&0.001&
                                             0.030&0.001\\
7065&
        0.0243&0.0002&
		0.0218&0.0002&&
			0.032&0.001&
				0.030&0.001&&
					0.025&0.001&
                                             0.026&0.001\\
7281&
        0.0073&0.0004&
		0.0066&0.0003&&
			0.006&0.001&
				0.006&0.001&&
					0.005&0.001&
                                             0.005&0.001\\
\enddata
\tablenotetext{a}{The $F$ values and their errors are those presented by the observers.
The $I$ values denote the intrinsic fluxes after correcting for underlying
absorption, collisional contribution to the Balmer lines, and interstellar
reddening; the errors attached to the $I$ values are only those due to the flux errors, see text.}
\tablenotetext{b}{$F$ values for region A from \citet{pei00}.}
\tablenotetext{c}{$F$ values from \citet{izo97}.}
\tablenotetext{d}{Including the contribution due to H8.}
\end{deluxetable}

\begin{deluxetable}{lr@{$\pm$}lr@{$\pm$}lcr@{$\pm$}lr@{$\pm$}l}
\tablecaption{Adopted He~{\sc{i}} line intensities relative to H$\beta$\tablenotemark{a}
\label{tab:lineB}}
\tablewidth{0pt}
\tablehead{
&
\multicolumn{3}{c}{SBS~0335-52\tablenotemark{b}}&&
\multicolumn{3}{c}{I~Zw~18\tablenotemark{c}}\\
\cline{2-5} \cline{7-10}  
\colhead{He~{\sc{i}} line}&
\multicolumn{2}{c}{$F$}&\multicolumn{2}{c}{$I$}&&
\multicolumn{2}{c}{$F$}&\multicolumn{2}{c}{$I$}}
\startdata

3889&
	0.1606&0.0018\tablenotemark{d}&
		0.0948&0.0026&&
			0.1570&0.0043\tablenotemark{d}&
                                0.0844&0.0052\\
4026&
	0.0122&0 0005&
		0.0179&0.0007&&
			0.0151&0.0036&
                              0.0222&0.0053\\
       										                                          
4471&
        0.0339&0.0007&
		0.0401&0.0008&&
			0.0352&0.0025&
                             0.0412&0.0029\\
4922&
        0.0077&0.0004&
		0.0094&0.0005&&
			\multicolumn{2}{c}{...}&
                              \multicolumn{2}{c}{...}\\
5876&
        0.1168&0.0014&
		0.1132&0.0014&&
			0.0968&0.0028&
                              0.1016&0.0029\\
6678&
        0.0322&0.0005&
		0.0297&0.0005&&
			0.0273&0.0019&
                              0.0290&0.0020\\
7065&
        0.0453&0.0007&
		0.0406&0.0006&&
			0.0249&0.0016&
                              0.0263&0.0017\\

\enddata
\tablenotetext{a}{The $F$ values and their errors are those presented by the observers.
The $I$ values denote the intrinsic fluxes after correcting for underlying
absorption, collisional contribution to the Balmer lines, and interstellar
reddening; the errors attached to the $I$ values are only those due to the observed 
flux errors, see text.}
\tablenotetext{b}{$F$ values for the three brightest positions by \citet{izo99}: center, 0".0SW., and 0".6NE.}
\tablenotetext{c}{$F$ values for the southeast region by \citet{izo99}.}
\tablenotetext{d}{Including the contribution due to H8.}
\end{deluxetable}
\clearpage

To correct for the stellar underlying 
absorption in the He{\sc~i} and H{\sc~i} lines we made use of synthetic models
and observations. 
The intensity of H$\beta$ was corrected taking into account the observed H$\beta$
equivalent widths in emission, $EW_{em }$(H$\beta$), and the H$\beta$ equivalent widths,
$EW_{ab }$(H$\beta$), based on the models by \citet{gon99,gon05} (Table~\ref{tab:Phys}).
The $EW_{ab }$ for the remaining Balmer lines and the He{\sc~i} lines were obtained
also from the models by \citet{gon99,gon05} and Cervi\~no (private communication), 
consistently with the adopted $EW_{ab }$(H$\beta$) values.
The He{\sc~i} $\lambda\lambda$ 7065 and 
7281 \AA\ lines are not included in the models by \citet{gon05}, therefore we made 
use of the observations by \citet{leo98} to correct for the stellar underlying
absorption of these lines.
 
After correcting for the underlying absorption and 
the collisional contribution to the Balmer lines presented in
Table~\ref{tab:all_collisions}, we determined the He{\sc~i} intensities
(due to recombinations and collisions, and affected by optical depth 
effects) relative to the pure recombination H$\beta$ line intensities 
uncorrected for reddening, $G$(He{\sc~i})/$G({\rm H}\beta)$.
Once these two corrections were made, we combined the 
$C({\rm H}\beta)^{\rm dust}$ value with the reddening law by Seaton 
(1979, hereafter S79) and derived the reddening corrected 
$I$(He{\sc~i})/$I({\rm H}\beta)$ line intensity ratios; note that 
$I({\rm H}\beta)$ represents the recombination contribution only, 
while the $I$(He{\sc~i}) values include recombination and collisional 
contributions as well as optical depth effects.

In Tables~\ref{tab:lineA} and \ref{tab:lineB} we present the 
$I$(He{\sc~i})/$I({\rm H}\beta)$ line intensity ratios and the 
errors associated with the observational measurements only; specifically, 
errors associated with underlying absorption, collisional contribution 
to the Balmer lines, and interstellar reddening are not included; they will 
be included in section 3.3 and discussed in section 3.4.
We decided to follow this procedure because the maximum likelihood method (MLM) 
used in section 3.2 to determine the helium physical conditions, requires the 
errors presented for each line to be independent from each other; while the 
effects of the uncertainties, from any one of these three sources 
(underlying absorption, collisional contribution to the Balmer lines, 
and interstellar reddening), affect the 
$I$(He{\sc~i})/$I({\rm H}\beta)$ ratios in a correlated way.

For SBS 0335-052 we used the observed $F$ values by \citet{izo99}. 
We did not use the data presented by \citet{izo06} for the brightest
1".56 x 1".04 region for this object because, even if the
signal to noise is very high, the observations were obtained
in seven different spectral ranges which did not correspond
exactly to the same region of the sky, this fact renders
these observations useless to determine a very accurate $Y$ value
\citep{izo06}. We also did not use the data for the whole object by
\citet{izo06} because the quality of the line intensity determinations
is considerably lower than the quality of the data presented by \citet{izo99}.

In Paper I the NGC 346 observations
were not corrected for underlying absorption in the He{\sc~i} and 
H{\sc~i} lines because in region A the bright O stars that ionize 
the H{\sc~ii} region were avoided. We consider that this decision 
was incorrect and that most of the continuous emission in region 
A is due to dust-scattered light showing the
He{\sc~i} and H{\sc~i} lines in absorption. The reasons are the following:
a) \citet{ode65}, and \citet{ode66} showed that most of the non-stellar continuum
observed in the visual range of the spectrum of H{\sc~ii} regions, when the
brightest stars are not included in the observing slit, is due
to dust-scattered light, b) comparing the equivalent width of region A, $EW$(H$\beta$) = 250 \AA,
to the expected $EW$(H$\beta$) for a low density plasma at 
12500 K, which amounts to $\sim$ 1230 \AA\ 
\citep[e.g.][and references therein]{all84,ost06}, 
it is found that about 80\% of the continuum light is not produced 
by the nebular gas, c) The color of the underlying spectrum obtained by subtracting the nebular continuum 
from the observed spectrum is bluer than that provided
by OB stars and therefore cannot be due to A and later type stars
(the intensity of the observed continuum at different wavelengths
can be obtained from the $EW$ of the emission lines presented in 
Table 5 of \citet{pei00}, and the absolute
line intensities presented in Table 2 of the same paper; the intensity of the nebular
continuum at different wavelengths can be obtained from Table 4-9 by \citet{all84} and the 
intensities at different wavelengths for stars of different spectral types 
can be obtained from Table 8 of the paper by \citet{cod60}).
Since the OB stars were avoided from 
the observations, it can neither be due to them.
We conclude from the previous arguments that most of the underlying 
continuum in region A of NGC 346 is due to dust-scattered light
provided by the brightest OB stars of the cluster.

\subsection{The $Y$ determinations}

The abundance analysis of these objects is based on the combined use 
of standard empirical relations and tailored photoionization
models. For NGC 2363, we used the models described
in \citet*{lur99}, while for NGC 346 we used the model by \citet*{rel02}.
As mentioned in Section~\ref{sec:collisions}, the models for the remaining 
three objects were presented in Paper II.

In addition to the collisional contribution to the Balmer lines, to
obtain He$^+$/H$^+$ values we need a set of effective recombination
coefficients for the helium and hydrogen lines, 
an estimate of the optical depth effects for the He{\sc~i} lines,
and the contribution to the He{\sc~i} line intensities  
due to collisional excitation.
We used the hydrogen recombination coefficients by \citet{sto95},
the helium recombination coefficients by \citet{por05}, with the 
interpolation formulae provided by \citet{por07}, and the 
collisional contribution to the He{\sc~i} lines by
\citet{saw93} and \citet{kin95}. The optical depth effects in the
triplet lines were estimated from the computations by \citet*{ben02}.

As in Paper I, we used a maximum likelihood method (MLM) based on the
He{\sc~i} line intensities 
to derive the $N$(He$^+$)/N(H$^+$) values. We produced two sets of models,
one where we assumed constant temperature $(t^2=0.000)$, given by $T$(4363/5007), and obtained
$\tau_{3889}$, $n_e$, and  $N$(He$^+$)/$N$(H$^+$), and the other set of models
where $T$(He$^+$) was an additional variable determined also by the MLM, in all cases $T$(He$^+$)
resulted smaller than $T$(4363/5007), to reconcile both temperatures it is necessary to
assume the presence of temperature variations characterized by $(t^2\ne0.000)$ (see 
\citet{pei67} and Paper I). The
resulting values are presented in Table~\ref{tab:Phys}, where the errors
include only those due to: the temperature structure, the density structure,
the optical depth of the He I triplet lines, and the adopted line intensities.

The MLM solutions with $t^2\ne0.000$ yield lower values of $N$(He$^+$)/$N$(H$^+$) than
those with $t^2=0.000$ (see Table~\ref{tab:Phys}). This change is not due to the 
variation of the He{\sc~i} recombination coefficients with temperature, which in this 
range of temperature is very small. The change is 
mainly due to the higher densities, derived from the MLM when $t^2\ne0.000$,
which increase the importance of the collisional excitation of the He{\sc~i} lines.
The densities for $t^2\ne0.000$ of the whole sample are about 67 cm$^{-3}$ higher
than for $t^2=0.000$. The MLM densities for NGC 346 and Haro~29 are $4 \pm 15$ and $-7 \pm 50$ cm$^{-3}$
respectively, these values are too low and support the idea that $t^2$ has to be larger than 0.000.
In Table~\ref{tab:Phys} we have adopted for NGC 346 and Haro~29 a density of 10 cm$^{-3}$ for $t^2=0.000$.
The rms density for NGC 346 is $14 \pm 3$ cm$^{-3}$ \citep{pei00}, a reasonable 
value for this type of objects. The rms density is the minimum density that can be associated to
an H{\sc~ii} region. For real nebulae, which always present large density variations, the local density
associated with the physical mechanisms that produce the line intensities, is always considerably higher than the rms 
density \citep[e.g.][]{ost59,pei66}.

The total helium to hydrogen abundance ratio was derived using the following
equation

\begin{eqnarray}
\frac{N ({\rm He})}{N ({\rm H})} & = &
\frac {\int{N_e N({\rm He}^0) dV} + \int{N_e N({\rm He}^+) dV} + 
\int{N_e N({\rm He}^{++})dV}}
{\int{N_e N({\rm H}^0) dV} + \int{N_e N({\rm H}^+) dV}},
						\nonumber \\
& = & ICF({\rm He})
\frac {\int{N_e N({\rm He}^+) dV} + \int{N_e N({\rm He}^{++}) dV}}
{\int{N_e N({\rm H}^+) dV}}
\label{eq:ICF}
,\end{eqnarray}
where $ICF$(He) is the helium ionization correction factor. The $ICF$(He) values 
were obtained from the Cloudy models and are presented in Table~\ref{tab:Phys}.
To obtain the $N$(He$^{++}$)/$N$(H$^+$) values, we used the $I$(4686)/$I$(H$\beta$)
value together with the recombination coefficients by \citet{bro71}.

\clearpage
\begin{deluxetable}{lccccc}
\rotate
\tablecaption{Physical Parameters for the H{\sc~ii} Regions
\label{tab:Phys}}
\tablewidth{0pt}
\tablehead{
\colhead{} &
\colhead{NGC 346} &
\colhead{NGC 2363} &
\colhead{Haro~29} &
\colhead{SBS 0335-052\tablenotemark{a}} &
\colhead{I Zw 18}}
\startdata
$EW_{em }$(H$\beta$)      & $250\pm10$      & $187\pm10$      & $224\pm10$       & $169\pm10$       & $135\pm10$       \\
$EW_{abs}$(H$\beta$)      & $2.0\pm0.5$     & $2.0\pm0.5$     & $2.0\pm0.5$      & $2.0\pm0.5$      & $2.9\pm0.5$      \\
$ICF$(He)                 & $1.000\pm0.001$ & $0.993\pm0.001$ & $0.9955\pm0.001$ & $0.991\pm0.001$  & $1.000\pm0.001$  \\[2.0ex]
$n_e(t^2=0.000)$          & $10\pm15$       & $205\pm67$      & $10\pm50$        & $230\pm41$       & $57\pm70$        \\
$\tau_{3889}(t^2=0.000)$  & $0.10\pm0.20$   & $1.12\pm0.40$   & $1.65\pm0.29$    & $2.65\pm0.35$    & $0.06\pm0.05$    \\
$N$(He$^+$)/$N$(H$^+$)$(t^2=0.000)$\tablenotemark{b}
                          & $8506\pm~48$    & $8570\pm175$    & $8651\pm162$     & $8542\pm146$     & $8451\pm354$     \\
$N$(He)/$N$(H)$(t^2=0.000)$\tablenotemark{b}
                          & $8528\pm~50$    & $8584\pm176$    & $8710\pm163$     & $8757\pm147$     & $8522\pm354$     \\
$N$(O)/$N$(H)$(t^2=0.000)$\tablenotemark{b}
                          & $11.5\pm1.8$    & $8.88\pm0.82$   & $7.43\pm0.75$    & $2.26\pm0.25$    & $1.66\pm0.17$    \\
$O$$(t^2=0.000)$\tablenotemark{c}   
                          & $139\pm 22$     & $107\pm 10$     & $~89 \pm ~9$     & $~27 \pm ~3$     & $~20 \pm~2$      \\[2.0ex]

$t^2$                     & $0.019\pm0.008$ & $0.021\pm0.011$ & $0.024\pm0.008$  & $0.040\pm0.010$  & $0.025\pm0.006$  \\
$n_e(t^2\ne0.000)$        & $58\pm33$       & $285\pm92$      & $61\pm50$        & $329\pm61$       & $90\pm80$       \\
$\tau_{3889}(t^2\ne0.000)$& $0.10\pm0.20$   & $1.04\pm0.40$   & $1.28\pm0.30$    & $2.56\pm0.35$    & $0.06\pm0.05$    \\
$N$(He$^+$)/$N$(H$^+$)$(t^2\ne0.000)$ \tablenotemark{b}
                          & $8372\pm~77$    & $8425\pm180$    & $8453\pm172$     & $8273\pm158$     & $8297\pm346$     \\
$N$(He)/$N$(H)$(t^2\ne0.000)$ \tablenotemark{b}
                          & $8399\pm~79$    & $8440\pm181$    & $8513\pm173$     & $8490\pm159$     & $8368\pm346$     \\
$N$(O)/$N$(H)$(t^2\ne0.000)$ \tablenotemark{b}
                          & $13.5\pm2.0$    & $10.7\pm0.8$    & $8.85\pm1.00$    & $3.26\pm0.33$    & $2.16\pm0.25$     \\
$O$$(t^2\ne0.000)$ \tablenotemark{c}      
                          & $163\pm 25$     & $129\pm 10$     & $106 \pm 12$     & $~39 \pm ~4$     & $~26 \pm ~3$     \\
\enddata
\tablenotetext{a}{Values for the three brightest positions by \citet{izo99}.}
\tablenotetext{b}{In units of $10^{-5}$.}
\tablenotetext{c}{Oxygen abundance by mass, in units of $10^{-5}$.}
\end{deluxetable}
\clearpage

{From} the normalization by unit 
mass given by $X + Y + Z = 1$, the $N$(He)/$N$(H) values, 
the $N$(O)/$N$(H) values, and the assumption that the oxygen by mass, 
$O$, amounts to 55\% $\pm 10\%$ of the $Z$ value, it is possible to derive 
$Y$ and $O$. The $N$(O)/$N$(H) and $O$ values are presented in Table~\ref{tab:Phys},
while the $Y$ values are presented in Table~\ref{tab:comp}.

The $O$/$Z$ = $0.55 \pm 0.10 $ value was derived by extrapolating to $O$ = 0 the 
$O/Z$ values for the Orion nebula, 30 Doradus in the LMC, and NGC 346 in the SMC 
derived by \citet{pea03}, which amount to $(43\pm 5)\% $, $(46\pm 7)\% $, and 
$(53 \pm 8)\% $, respectively.
Note that NGC 346 is the $O$ richest object in our sample and that the increase in
the $O$ fraction with decreasing $Z$ is mainly due to the decrease of the $C/O$ 
and $N/O$ ratios with decreasing
$O$ abundance. The error in the $O/Z$ ratio translates into an error slightly smaller than
0.0001 in the $Y_p$ determination.

The $O$ values in Table~\ref{tab:Phys} are slightly different to those presented
in Paper I due to four causes: a) the $N$(O)/$N$(H) values are higher
because we took into account the collisional contribution to the Balmer line
intensities, b) the $N$(He)/$N$(H) ratios are higher, c) the $t^2$ values are 
slightly different, because the adopted
line intensities are slightly different, and d) we assumed that 10\% of the
O atoms are trapped in dust grains in all objects.

In Table~\ref{tab:comp} we explicitly present the $\Delta Y$ increase due to the
collisional excitation of the Balmer lines. Also in this table we present the $Y$ values
for $t^2 = 0.000$ and those derived for $t^2\ne0.000$. The  $Y$ and $Y_p$ values present
first the statistical and then the systematic errors from all the sources presented in Table~\ref{tab:error} 
(note that the $N$(He)/$N$(H) ratios presented in  
Table~\ref{tab:Phys} only include a subset of the statistical errors, those due to: temperature structure, density structure, optical depth effects and line intensities). For each object we determine a $Y_p$ value and at the 
end of the table we present the $Y_p$ value for the whole
sample, $Y_p$(sample). The error budget for the $Y_p$(sample) is discussed in the
error budget subsection.

\subsection{The $Y_p$ determination}

To determine the $Y_p$ value from all the objects it is necessary 
to estimate the fraction of helium present in the interstellar 
medium produced by galactic chemical evolution. We will assume that
\begin{equation}
Y_p  =  Y - { O} \frac{\Delta Y}{\Delta O}
\label{eq:DeltaO}
,\end{equation}
where $O$ is the oxygen abundance by mass.  From chemical evolution 
models of different galaxies it is found that $\Delta Y/\Delta O$
depends on the initial mass function, the star formation rate, 
the age, and the $O$ value of the galaxy in question. But $\Delta Y/\Delta O$ is well fitted by a 
constant value for objects with the same IMF, the same age, and 
an $O$ abundance smaller than $\sim$ 4$\times 10^{-3}$ \citep{pei07}.
Consequently in what follows we will adopt a constant value 
for $\Delta Y/\Delta O$ .

The $\Delta Y/\Delta O $ value derived by \citet{pei00} from observational
results and models of chemical evolution of galaxies amounts to
3.5 $\pm$ 0.9. More recent results are those by \citet{pea03}, who 
finds 2.93 $\pm$ 0.85 from observations
of 30 Dor and NGC 346, and by \citet{izo06} who  find 
$\Delta Y/\Delta O $ = 4.3 $\pm~0.7$ from the observations
of 82 H{\sc~ii} regions. We
have recomputed the Izotov et al. value considering two systematic effects
not considered by them: the 
fraction of oxygen trapped in dust grains, which we estimate to be 10\%, and 
the increase in the inferred $O$ abundances due to the presence of temperature 
fluctuations, which for this type of H{\sc~ii} regions we estimate to be 
about 0.08 dex \citep{rel02}. {From} these considerations we obtain 
$\Delta Y/\Delta O $ = 3.2 $\pm 0.7$. On the
other hand \citet{pei07} from chemical evolution models with
different histories of galactic inflows and outflows for objects 
with $O < 4 \times 10^{-3}$ find that 
2.4 $< \Delta Y/\Delta O < 4.0$. {From} the theoretical and observational
results we have adopted a value of $\Delta Y/\Delta O $ = 3.3 $\pm$ 0.7, which 
we have used with the $Y$ and $O$ determinations from each object to 
obtain the set of $Y_p$ determinations presented in Table ~\ref{tab:comp}.

To determine the $Y_p$ average from the whole sample we first need to find the weight 
that should be assigned to each object by considering the confidence we have in each one
of the determinations. For this we added in quadrature the errors provided by the MLM, presented in 
Table~\ref{tab:Phys}, plus all the additional sources of error presented in Table ~\ref{tab:error}, with the exception of the errors associated with the recombination 
coefficients of both H and He (which will affect the sample as a whole and thus 
will not affect the relative confidence we have in the determination from each object). 
The quadratic addition of these 11 sources of error result in: 
$err_{\Sigma 11}$(NGC~346)=0.00311, $err_{\Sigma 11}$(NGC~2363)=0.00507, 
$err_{\Sigma 11}$(Haro~29)=0.00475, $err_{\Sigma 11}$(SBS~0335-052)=0.00583, 
and $err_{\Sigma 11}$(I~Zw~18)=0.00867.
And the weights obtained in this way, for each object in the sample, 
amount to: $w$(NGC~346)=0.4515, 
$w$(NGC~2363)=0.1697, $w$(Haro~29)=0.1927, 
$w$(SBS~0335-052)=0.1281, and $w$(I~Zw~18)=0.0579.

We use these weights to determine the different helium averages 
$\left<Y\right>=\sum_i Y(i) w(i)$ along with the corresponding statistical errors
$err_{sta}=\bigl(\sum_i [err_{sta}(i) w(i)]^2\bigr)^{1/2}$ and systematic errors 
$err_{sys}=\sum_i err_{sys}(i) w(i)$; finally we add the statistical and systematic 
errors in quadrature.

To compare our results with those of other authors that assume $t^2 = 0.000$ we have 
computed $Y_p(t^2=0.000)= 0.2523 \pm 0.0027$, using the five $Y$ and $O$ values for $t^2 = 0.000$ 
presented in Tables \ref{tab:Phys} and~\ref{tab:comp}, note that this value is not presented in 
Table~\ref{tab:comp}. Also, from the five $Y_p$ values for $t^2 \neq 0.000$ presented in 
Table~\ref{tab:comp}, we derive $Y_p = 0.2477 \pm 0.0029$.

\clearpage
\begin{deluxetable}{lcccc}
\rotate
\tablecaption{$Y$ and $Y_p$ values
\label{tab:comp}}
\tablewidth{0pt}
\tablehead{ &&\colhead{$Y$} &\colhead{$Y$} &\colhead{$Y_p$\tablenotemark{a}}\\ 
&\colhead{$\Delta Y$(Hc)\tablenotemark{b}} & \colhead{$t^2=0.000$} & \colhead{$t^2 \neq 0.000$} & \colhead{$t^2 \neq 0.000$}} 
\startdata
NGC~346           &$0.0015\pm0.0005$ & $0.2537 $ & $0.2507 \pm 0.0027 \pm 0.0015$ & $0.2453 \pm 0.0027 \pm 0.0019$ \\
NGC~2363          &$0.0057\pm0.0016$ & $0.2551 $ & $0.2518 \pm 0.0047 \pm 0.0020$ & $0.2476 \pm 0.0047 \pm 0.0022$ \\
Haro~29           &$0.0047\pm0.0013$ & $0.2577 $ & $0.2535 \pm 0.0045 \pm 0.0017$ & $0.2500 \pm 0.0045 \pm 0.0019$ \\
SBS~0335--052     &$0.0144\pm0.0038$ & $0.2594 $ & $0.2533 \pm 0.0042 \pm 0.0042$ & $0.2520 \pm 0.0042 \pm 0.0042$ \\
I~Zw~18           &$0.0114\pm0.0031$ & $0.2529 $ & $0.2505 \pm 0.0081 \pm 0.0033$ & $0.2498 \pm 0.0081 \pm 0.0033$ \\[2.0ex]
Sample            &$0.0056\pm0.0015$ & $0.2554 $ & $0.2517 \pm 0.0018 \pm 0.0021$ & $0.2477 \pm 0.0018 \pm 0.0023$\tablenotemark{c} \\ 
\enddata
\tablenotetext{a}{Derived from each object under the assumption that 
$\Delta Y/\Delta O = 3.3 \pm 0.7$ see text.}
\tablenotetext{b}{Increase in the $Y$ abundance due to the collisional contribution to the
Balmer line intensities.}
\tablenotetext{c}{Equal to $0.2477 \pm 0.0029$.}
\end{deluxetable}
\clearpage

\subsection{The error budget}\label{sec:error}

Based on the errors presented in Table~\ref{tab:Phys} and Table~\ref{tab:comp}, and
the discussion in this subsection, 
we have elaborated the error budget 
for the whole sample of our $Y_p$ determination and it is presented in 
Table~\ref{tab:error}. The errors in the table are grouped in discrete bins because they represent
broad estimates. In this table the sources of error are listed in order
of importance, we will say a few words for some of them. The
error budgets of other $Y_p$ determinations are different to ours for
many reasons, they depend on the sample of H{\sc~ii} regions and on the
treatment given by the different groups to each source of error.

It is difficult to estimate the magnitude of the $Y_p$ error caused by
each of the thirteen sources of error, since there is a significant
interdependence between some of them; for instance, the uncertainty in
the hydrogen collisions will modify the reddening determination for
each object. To quantify the total effect of each source of error in
the $Y_p$ determination we used the following approach: a) we started
with the four errors derived from the MLM (He~{\sc{i}} and H~{\sc{i}}
line intensities, optical depth of the He~{\sc{i}} triplet lines,
collisional excitation of the He~{\sc{i}} lines due to the average
density, and temperature structure); b) we ordered the other sources
of error to be considered one at a time (in order: density structure,
helium $ICF$, underlying absorption of the H~{\sc{i}} and He~{\sc{i}}
lines, reddening correction, collisional excitation of the H~{\sc{i}}
lines, $O$ $(\Delta Y/\Delta O)$ correction, and recombination
coefficients of the H~{\sc{i}} and He~{\sc{i}} lines); c) we measured
how a change in each source affects the determination of all the
previous quantities; d) we presented the total amount as the estimated
contribution to the $Y_p$ error for this latest source, thus all the
cross correlations with the previous sources of error are included
in the later source. For example, the $Y_p$ error produced by the
modification of the reddening correction due to the hydrogen
collisions is presented as part of the error due to the hydrogen
collisions, and is not presented as part of the error due to the
reddening correction; also not presented, is the increase in the
$C({\rm H}\beta)$ formal error due to the uncertainty of the hydrogen
collisions determinations.

Estimated in this manner each source of error is independent from the
others, even the ones labeled as systematic. Consequently the standard
deviation of the total error is computed by adding in quadrature the
standard deviation of all the sources of error. Note that we are not
assuming that any of the thirteen errors has a normal distribution
(even if several of the error distributions probably are normal), we
are reporting what we estimate to be the standard deviation of the
$Y_p$ error due to each source; also, with thirteen independent
sources of error included, we expect the final error to be close to a
normal distribution.

In Table~\ref{tab:error} we have divided the sources of error in two
groups, statistical and systematic. The errors labeled as statistical
will affect differently the $Y_p$ determination for each object;
therefore, increasing the number of objects in the sample will reduce
their final magnitude. On the other hand the errors labeled as
systematic, are so in the sense that for any one of them our lack of
understanding shifts the $Y_p$ determination of each object in the
same direction, for some of them by different amounts, producing an
error in the determination that cannot be reduced by simply increasing
the number of objects; for some of them the error can be diminished by
selecting a small group of objects where we expect this effect to be
minimum.  The specific equations that should be used to determine the
errors from a sample, which differentiate the treatment given to the
systematic and statistical errors, are presented in the fourth
paragraph of section 3.3 and were used to estimate the final error of
the $Y_p$ determination.  

As expected the total error obtained from Table~\ref{tab:error} is in
agreement with the total error presented in Table~\ref{tab:comp} and
implies that the estimates of the different sources of errors in both
tables are equivalent.

The most important source of error in the determination of $Y_p$ is the collisional 
excitation of the Balmer lines. The collisional contribution
to the Balmer lines produces a non-negligible increase in the Balmer
line intensities relative to the case B recombination and could 
introduce a bias in the reddening correction deduced from the observed H$\alpha$/H$\beta$
ratio, both effects affecting the $Y$
determination. For our sample the Balmer collisional contribution on the reddening correction
is practically negligible (see Table~\ref {tab:all_cHbeta}). On the other hand the 
collisional contribution to the Balmer line intensities is very important in the $Y$ determination
(see Table~\ref {tab:comp}). We will give an approximate estimate of the error introduced by this
effect on our $Y_p$ determination. We will consider three 
sources of error: the collisional rates, the radiative cascade, and the model fitting (see section~\ref {sec:collisions}). The published collisional rates are expected to  become increasingly accurate with time, and the current uncertainty  on them, estimated from the variation of the $\Omega$s published in  the last decade, is of the order of 15\% for $\Omega(1s, 3)$ and $\Omega(1s, 4)$ in the temperature range of interest to us. These values translate into almost identical values for $I$(H$\alpha_\mathrm {col}$) and 
$I$(H$\beta_\mathrm{col}$), which in turn would  correspond to an uncertainty of about 0.0009 in $Y_p$.
The errors on the assumed path for the radiative cascade are  negligible, since the breakdown 
of a given $\Omega(1s,n)$ among  sublevels is much less uncertain than the absolute value 
of the $ \Omega(1s, n)$ itself. On the other hand, model fitting of the observed  H{\sc~ii} region is the leading source of  uncertainty, because the ionization and temperature structures are  difficult to constrain to the high level of accuracy that would be  desirable given the strong dependence of the collisional rates on the  electron temperature and the neutral hydrogen fraction. In Section~\ref {sec:collisions}, we have shown that the collisional contribution to  the Balmer lines mostly comes from the region where the [O II] lines  originate, so that it is important for a good model to reproduce the observed $T$[O II] temperature. Our estimate of the uncertainty is based on the  range covered by the collisional contribution in models that acceptably fit the $T$[O II] temperature. 
By comparing reasonable photoionization models for the same object we consider that the model fitting
introduces an error of about 22\% on the collisional contribution to the Balmer lines. The
combination of the three sources of error amounts to about
27\% of the collisional contribution to the Balmer lines of our sample, which translates into an error of 
about 0.0015 in $Y_p$.

We consider the error associated
with the collisional excitation of the H~{\sc{i}} lines as systematic because
this process has not been studied at length by different groups, hence this effect could be systematically
lower or higher for all the objects. The error 
might become statistical once the problem is studied further, but in any case
we expect it to depend strongly on the photoionization models needed for its estimation.

The second most important source of error is the temperature structure. Most $Y$
determinations are based on $T$(4363/5007), but other temperature determinations
yield lower values, and photoionization models do not predict the high
$T$(4363/5007) values observed. These results indicate the presence
of temperature variations which should be included in the $Y$ determination
(see Paper I). The best procedure to take into account the temperature
structure is to determine 
$T$(He$^+$) based on the maximum likelihood method. The $Y$ abundances derived 
from $T$(He$^+$) are typically lower by about 0.0040 than those derived from $T$(4363/5007).
The error quoted in Table~\ref{tab:error} is due to the error in the $T$(He$^+$)
determinations obtained with the MLM (see section 3.2).
The difference between both temperatures is not correlated to the metallicity
of the H{\sc~ii} region,
therefore the systematic error introduced by the use of $T$(4363/5007) 
in the $Y$ determination is similar for objects with different metallicities.

We have estimated the error in the $Y_p$ determination due to the adopted
effective recombination coefficients based on the confidence in the He~{\sc{i}} line
emissivities presented by \citet{bau05}. The lines used to determine the helium
abundances are $\lambda\lambda$ 3820A, 3889B, 4026AA, 4387A, 4471A, 4921A, 5876A,
6678A, 7065A, and 7281A, where the letter indicates the confidence: AA better than 0.1\%,
A in the 0.1 to 1\% range, and B in the 1 to 5\% range. From these values we estimate
a systematic error due to the computed emissivities of about 0.0010 in $Y$. According 
to \citet{por07}, the error introduced in the emissivities 
by interpolating the equations provided by them in temperature is smaller than
0.03\%, which translates into an error in $Y_p$ considerably smaller than 0.0001.
In our preliminary estimate of $Y_p$ \citep{pei07} we used a different interpolation 
to the \citet{por05} He{\sc~i} atomic recombination coefficients than that given
by \citet{por07}, as well as a slightly different error budget, and obtained 
that $Y_p = 0.2474 \pm 0.0028$.

The third most important source of error is the extrapolation to
zero heavy elements content. Fortunately, based on chemical evolution models
of galaxies of different types, it is found that $\Delta Y/\Delta O $
is practically constant for objects with $O < 4 \times 10^{-3}$,  
and in good agreement with the observational determinations (see the
previous section). One of the sources of error in the observational 
determination is the fraction of oxygen trapped in dust grains 
which has to be taken into account.

As was demonstrated by \citet{oli04}, the reddening correction is an important 
source of error. Therefore to estimate the error in the reddening correction
we will make comparisons among three classic extinction laws and a recent one, 
these laws are labeled: S79, W58 \citep[Whitford 1958; as parameterized by] []{mil72}, 
CCM89 \citep{car89}, and B07 \citep{bla07}. Fortunately,
for our sample the average $C$(H$\beta$) value amounts to 0.09 only,
see Table~\ref{tab:all_cHbeta}. For the 4500 to 7300 \AA\ region the differences
among the S79, W58 and CCM89
extinction laws produce differences in the $I$(He$\lambda$)/$I$(H$\beta$) 
values of our sample of about 0.15\%, which correspond to an average
difference smaller than 0.0002 in the final $Y_p$ determination. For the 
3800 to 4400 \AA\ region the differences between S79 and W58 produce differences
of about 0.4\% in the line intensity ratios, which, when combined with all the  He~{\sc{i}} lines
available, correspond to differences smaller than 0.0004 on $Y_p$. For the 
3800 to 4400 \AA\ region the use of the CCM89 law for $R_V$ = 3.16, instead of the S79
law used by us, gives systematically higher $I$(He$\lambda$)/$I$(H$\beta$)
values by about 1\%, increasing $Y_p$ by about 0.0008 relative to our result when all the He I lines presented
in Table~\ref{tab:lineA} and Table~\ref{tab:lineB} are used. \citet{car89} made a strong effort to find a simple
analytical law to be useful for all $R_V$ values, we consider that in the 3800 to 4400 \AA\
range the CCM89 law might overestimate the extinction because it might not be perfectly
represented by a seventh order polynomial. Support for this idea comes from \citet{bla07},
who find that the CCM89 law for $R_V=5.5$ overestimates the extinction in the 
3030 to 4350\AA\ region when compared with the B07 law. The B07 extinction law is intended to 
reproduce the Orion reddening law using a modified CCM89 extinction ($R_V=5.5$) 
which only differs from the CCM89 one in the 3030 to 4350\AA\ region. Based on the 
previous discussion we estimate that the error in the $Y_p$ determined by us due to the reddening
correction amounts to about 0.0007 and we consider that most of this error is systematic. Note that
for a sample with an average $C$(H$\beta$) of 0.2 our estimate of the error in $Y_p$
increases to about 0.0016.

It has often been shown that the correction for underlying absorption in the H~{\sc{i}} and He~{\sc{i}} lines has been underestimated \citep{ski98,oli04,por07}. We consider that the best procedure to correct for
underlying absorption is to use the models by \citet{gon99,gon05}. According to these models
for those objects with
$EW_{em}$(H$\beta$) $\ge 150$\AA\ the expected $EW_{ab}$(H$\beta$) amounts to $\approx 2$\AA\ ;
for objects with $EW_{em}$(H$\beta$) $\le 150$\AA\, (older objects), the expected $EW_{ab}$(H$\beta$)
becomes larger. Therefore the correction for underlying absorption for objects with
$EW_{em}$(H$\beta$) $\ge 150$\AA\ is inversely proportional to $EW_{em}$(H$\beta$), while
for objects with $EW_{em}$(H$\beta$) $\le 150$\AA\ the correction, and consequently the associated error,
increases even faster due to the larger $EW_{ab}$(H$\beta$) predicted by the models.
For samples including a large fraction of objects with $EW_{em}$(H$\beta$) $\le 150$\AA\ we expect the error
for this concept to be larger than ours, moreover to agree with the models by
\citet{gon99,gon05} the average $EW_{ab}$(H$\beta$) for any given sample has to be equal or
larger than 2\AA\ .

In principle, the larger the number of He{\sc~i} lines used to 
determine a given $Y$ value the better. But there are two additional 
issues that have to be considered for some of the He{\sc~i} lines that are not 
included in our determination: a) the accuracy of the atomic
data for some of the lines is lower than that for those lines which we have used 
\citep{bau05}, and b) the radiative transfer of the He{\sc~i} singlet lines
has to be taken into account for the $p-s$ and $s-p$ transitions 
\citep{rob73,rob74}.

It is possible that further work might uncover additional sources of systematical
errors that would increase the final
error. Increasing the sample with objects as well observed as those in our sample
and with tailor made photoionization models of similar quality as those used by us will reduce
the statistical errors but not the systematic ones.
It is also possible that some of
the errors that we consider to be dominated by systematics might be
reduced by further work, for example by choosing a different set of objects where a particular
systematic effect is expected to be lower, or by
providing more detailed observations, or by increasing the accuracy of the atomic data
determinations, or by providing more realistic models for the observed H II regions.

\clearpage
\begin{deluxetable}{lcc@{$\pm$}r@{}c}
\tablecaption{Error budget in the $Y_p$(sample) determination
\label{tab:error}}
\tablewidth{0pt}
\tablehead{
\colhead{Problem} & \multicolumn{4}{c}{Estimated error}}
\startdata
Collisional Excitation of the H~{\sc{i}} Lines   &\hfill&& 0.0015\tablenotemark{b} &\hfill\\
Temperature Structure                                  &&& 0.0010\tablenotemark{a} &\\
$O$ $(\Delta Y/\Delta O)$ Correction                   &&& 0.0010\tablenotemark{b} &\\
Recombination Coefficients of the He~{\sc{i}} Lines    &&& 0.0010\tablenotemark{b} &\\
Collisional Excitation of the He~{\sc{i}} Lines        &&& 0.0007\tablenotemark{a} &\\
Underlying Absorption in the He~{\sc{i}} Lines         &&& 0.0007\tablenotemark{a} &\\
Reddening correction                                   &&& 0.0007\tablenotemark{b} &\\
Recombination Coefficients of the H~{\sc{i}} Lines     &&& 0.0005\tablenotemark{b} &\\
Underlying Absorption in the H~{\sc{i}} Lines          &&& 0.0005\tablenotemark{a} &\\
Helium Ionization Correction Factor                    &&& 0.0005\tablenotemark{a} &\\
Density Structure                                      &&& 0.0005\tablenotemark{a} &\\
Optical Depth of the He~{\sc{i}} Triplet Lines         &&& 0.0005\tablenotemark{a} &\\
He~{\sc{i}} and H~{\sc{i}} Line Intensities            &&& 0.0005\tablenotemark{a} &\\
\enddata
\tablenotetext{a}{Statistical error.}
\tablenotetext{b}{Systematic error.}
\end{deluxetable}
\clearpage

\section{Comparison with other $Y_p$ determinations}\label{sec:comparison}

The difference between our $Y_p$ value of 0.2477 and the 0.2391
value presented in Paper II  amounts to 0.0086, and is mainly due
to: the change in the He{\sc~i} recombination coefficients (which produced an increase in 
$Y_p$ of about 0.0040), the change in the 
H{\sc~i} collisional excitation coefficients (which produced an increase 
in $Y_p$ of about 0.0025), the correction for underlying absorption in 
the red He{\sc~i} lines, and
the correction of NGC 346 for underlying absorption. In Paper II we 
used different He{\sc~i} recombination coefficients, 
those by \citet{smi96} and \citet*{ben99}.

It is beyond the scope of this paper to present an error budget
for the $Y_p$ determinations of other authors, but we will discuss
some of the reasons responsible for the different $Y_p$ values
derived by other authors.

There are many differences with respect to the procedure followed by \citet{izo04},
who derived $Y_p = 0.2421 \pm 0.0021$.
At least two of them are systematic: our use of the 
recent He{\sc~i} recombination coefficients by \citet{por05,por07}, which yield $Y$ values 
about 0.0040 higher than the previous ones, and our use of the recent 
H{\sc~i} collisional data (see Section~\ref{sec:collisions}), 
which further increase the $Y$ values over the older H{\sc~i} collisional 
corrections by about 0.0025.

\citet{oli04} find a $Y_p$ = $0.249 \pm 0.009$. Our result is 
in agreement with theirs, but our error is smaller. Again there are 
the systematic differences due to the He{\sc~i} recombination data used by both groups
and to the estimation of the collisional contribution to the H Balmer lines, 
these two effects probably would increase their result by about 0.006. 

\citet{fuk06} based on a reanalysis of the \citet{izo04}
sample of 33 H{\sc~ii} regions determined a value of $Y_p$ = $0.250 \pm 0.004$. 
In addition to a different treatment of the underlying H and He{\sc~i} 
absorption there are four systematic effects between their
determination and ours. \citet{fuk06} used the He{\sc~i} recombination data by 
\citet{ben99}, did not take into account the
collisional excitation of the Balmer lines, adopted the 
$T$(4363/5007) value instead of the temperature provided 
by the He{\sc~i} lines to determine $Y$, and
assumed that $ICF$(He) = 1.000. Including 
the first two effects would increase their $Y_p$ value by
about 0.009, while the consideration of the third effect
decreases their determination by about 0.004, the assumption
of $ICF$(He) = 1.000 has to be tested with photoionization
models, for our sample three of our models showed $ICF$(He) values
smaller than 1.000. By using our sample as representative of their sample
(which might not be true) their $Y_p$ value gets reduced by 0.001. 
Another problem with their determination is that it implies a decrease of
the He{\sc~i} underlying absorption with metallicity, which is not
expected; what is expected instead is a decrease of the underlying
absorption with an increase of the equivalent width in emission
of H$\beta$.

\section{Cosmological implications}\label{sec:cosmology}

To compare our $Y_p$ value with the primordial deuterium abundance
$D_p$ (usually expressed as 10$^5({\rm D}/{\rm H})_p$) and with the WMAP 
results, we will use
the framework of the standard big bang nucleosynthesis. The ratio 
of baryons to photons multiplied by 10$^{10}$, $\eta_{10}$, is 
given by \citep{ste06b}:
\begin{equation}
\eta_{10} = (273.9 \pm 0.3)\Omega_bh^2,
\end{equation}
where  $\Omega_b$ is the baryon closure
parameter, and $h$ is the Hubble parameter. In the range 
$4\lesssim \eta_{10} \lesssim 8$ (corresponding to 
$0.2448\lesssim Y_p \lesssim 0.2512$), $Y_p$ is related to
$\eta_{10}$ by \citep{ste06a}:
\begin{equation}
Y_p = 0.2384 + \eta_{10}/625.
\end{equation}
In the same $\eta_{10}$ range, the primordial deuterium abundance is given by
\citep{ste06a}:
\begin{equation}
10^5({\rm D}/{\rm H})_p = D_p = 46.5\eta^{-1.6}_{10}.
\end{equation}
From our $Y_p$ value, the $D_p$ value by \citet{ome06},
the $\Omega_bh^2$ value by \citet{spe06}, and the previous
equations we have produced Table~\ref{tab:BBN}.
From this table, it follows that within the errors
 the $Y_p$, $D_p$, and $WMAP$ observations 
are in very good agreement with the predicted SBBN values. 

\clearpage
\begin{deluxetable}{lcccc}
\tablecaption{Cosmological predictions based on SBBN and observations.
\label{tab:BBN}}
\tablewidth{0pt}
\tablehead{
\colhead{Method}
&\colhead{$Y_p$}
&\colhead{$D_p$}
&\colhead{$\eta_{10}$}
&\colhead{$\Omega_bh^2$}}

\startdata
$Y_p$&
	$0.2477\pm0.0029$\tablenotemark{a}&
		$2.78^{+2.28}_{-0.98}$\tablenotemark{b}&
			$5.813\pm1.81$\tablenotemark{b}&
				$0.02122\pm0.00663$\tablenotemark{b}\\
					
$D_p$&
	$0.2476\pm0.0006$\tablenotemark{b}&
		$2.82\pm0.28$\tablenotemark{a}&
			$5.764\pm0.360$\tablenotemark{b}&
				$0.02104\pm0.00132$\tablenotemark{b}\\
					
$WMAP$&
	$0.2482\pm0.0004$\tablenotemark{b}&
		$2.57\pm0.15$\tablenotemark{b}&
			$6.116\pm0.223$\tablenotemark{b}&
				$0.02233\pm0.00082$\tablenotemark{a}\\

\enddata
\tablenotetext{a}{Observed value.}
\tablenotetext{b}{Predicted value.}
\end{deluxetable}
\clearpage

\section{Discussion}\label{sec:discussion}

The effect of collisions on the H Balmer spectrum emerges from this work as  
the leading source of uncertainty for our sample. 
This uncertainty has in turn three independent sources:
the theoretical uncertainty on the collisional $\Omega$s, 
the incompleteness of the collisional $\Omega$s, and the uncertainty on
the physical conditions of the gas at which collisions occur. 

The last of these factors is probably the most severe.
The collisional excitation of the Balmer lines is stronger in the hotter zones of the
H{\sc~ii} regions (which are predicted to be the innermost in this metallicity range) 
and in the less ionized zones (which are the outermost). Since the fraction of neutral hydrogen 
in a nebula varies over a much wider range than the Boltzmann factor, which is the main temperature dependence
of collisions, the contribution of the outer zones dominates, so that $T$(O{\sc~ii}) is the most appropriate 
temperature to characterize collisions when trying to model them. 
Additionally, in observations covering a large fraction of the nebula the 
emitting volume of the outer parts outweighs that of the inner parts, strengthening this conclusion.
However, H{\sc~ii} regions that are density bounded or have a strongly inhomogeneous temperature structure might escape this
rule, particularly when observed with small apertures. In these cases the uncertainty will be larger.

It is clear that only detailed observations and tailored modeling will allow us to reduce the uncertainty
introduced by the collisional excitation of the Balmer lines, particularly 
in the extreme cases; but a better choice might be to avoid critical H{\sc~ii} regions altogether, by preferentially selecting those H{\sc~ii} regions in which collisions are known in advance to play only a minor role, i.e. objects with moderate temperatures. 
Since the temperature in H{\sc~ii} regions is mostly determined by metallicity, this amounts to saying that 
metal-poor objects
are more adequate than extremely metal-poor objects in $Y_P$ determinations.
This conclusion runs counter the common wisdom that the
best candidates for primordial helium determinations 
are extremely metal-poor objects ($Z \lesssim 0.0005$),
because they permit to minimize the error introduced by
the extrapolation of the ($Y, O$) relation to 
$\mathrm{O/H}=0$: although this is true, a larger uncertainty on $Y_p$ 
is introduced by collisions than the one introduced by most of the other sources, 
including the slope $\Delta Y/\Delta O$,
so the observational efforts should be better directed at metal-poor objects.

Further disadvantages of extremely metal-poor objects in the quest for primordial helium
are that their number is very small and that their H{\sc~ii} regions are relatively faint.
These disadvantages, together with the uncertainty on collisions discussed above, 
largely outweigh the advantage implied by 
the smaller error introduced by the extrapolation of the ($Y, O$) relation to 
zero metallicity. Therefore, we are led to the strong conclusion 
that not so extremely metal poor objects, like those in the $0.0005\lesssim Z \lesssim 0.001$ range,
are more appropriate for the determination of $Y_p$:
in this range of metallicity it is possible to find a large set of 
objects with bright H{\sc~ii} regions, which will improve the 
quality and number of emission lines available for the determination 
of physical conditions; furthermore, the temperatures of these objects are
smaller than those of more metal-poor objects and consequently the 
correction due to the collisional excitation of the H~{\sc{i}} lines 
is also smaller.

This conclusion leads us to another critical point in the approach to $Y_p$.
It has been noted that the uncertainty on $Y_p$ is dominated by systematic errors
\citep{oli04}. As long as this is the case, it is preferable 
to analyze a few objects in depth and try to correct for the systematics
than to perform a more shallow analysis of a larger sample, since this last method can reduce 
the statistical uncertainties but not the systematic errors. 
Hence we strongly support the methodological choice
of understanding as well as possible the details of the objects in a small sample, 
before directing our efforts towards extending the sample. 
It is only by means of this approach that systematical errors, 
such as those arising from the temperature structure of H{\sc~ii} regions, can be gradually
understood, corrected for, and eventually transformed into statistical uncertainties.

\section{Conclusions}\label{sec:conclusions}

The new He{\sc~i} atomic recombination coefficients by \citet{por05,por07}
increase the $Y_p$ determination by about 0.0040 with respect to Paper II.

The H collisional excitation coefficients by \citet{and00,and02}
increase the $Y_p$ determination by about 0.0025 with respect to Paper II.

The adoption of temperature variations, instead of the 
assumption of constant temperature given by $T$(4363/5007),
reduces the $Y_p$ determinations by about 0.003 $-$ 0.006. In this paper
and in Paper II we did take into account temperature variations and
therefore there is no systematic difference due to this effect
between our two $Y_p$ determinations. On the other hand, when
comparing our results with those of other authors who assume a
constant temperature given by $T$(4363/5007), a systematic
difference is present that has to be taken into account.

From the analysis of the data studied in this paper we derive a $Y_p$ value of $0.2477 \pm 0.0029$
for $t^2 \neq 0.000$, while under the assumption of $t^2 = 0.000 $, we
obtain that $Y_p = 0.2523 \pm 0.0027$. From the $Y_p$ given by the WMAP results combined with
the adoption of the SBBN, if our error budget is correct, our result implies that $t^2$ 
has to be larger than zero at more than the 1-$\sigma$ level.

It has been argued that the uncertainty on $Y_p$ is dominated by systematic errors
\citep{oli04}. If this is the case, it is certainly preferable to analyze a few objects in depth
and try to correct for the systematical errors, than to rely on a statistical analysis 
of a large sample, since this last method can reduce the statistical errors
but not the systematic ones.

From Table~\ref{tab:BBN} it follows that the $Y_p$, $D_p$, and $WMAP$ 
observations are in very good agreement with the SBBN predicted values.
Moreover our $Y_p$ value, if our error budget is correct, provides stronger 
constraints to some predictions of non-standard Big Bang cosmologies than previous studies
\citep[e.g.][and references therein]{cyb05,coc06}.

\acknowledgments

The authors wish to thank Miguel Cervi\~no for help with the high
resolution synthesis models and Gary Ferland for several discussions
on this topic.  We are grateful to the referee for the detailed
reading of the manuscript and many excellent comments and suggestions.
We also acknowledge fruitful correspondence with R. L. Porter, K. B.
MacAdam, and G. Steigman.  The high-resolution synthesis models used
in this work have been retrieved from PGos3
(http://ov.inaoep.mx/pgos3), which is operated by the future Mexican
Virtual Observatory.  This work was partly supported by the CONACyT
grant 46904 and by the Spanish {\it Programa Nacional de Astronom\'\i
  a y Astrof\'\i sica} through projects AYA2004-02703 and
AYA2004-07466. VL acknowledges the hospitality of IA-UNAM and INAOE,
where a part of this research was carried out, and the support of a
{\it CSIC-I3P} fellowship.

\end{document}